\documentclass[runningheads]{llncs}
\usepackage{tabularray}
\usepackage{graphicx}
\usepackage{amsmath}
 \usepackage{url,amsfonts,amssymb,booktabs,makecell}
\begin{document}
\title{Effective Audio Classification Network Based on Paired Inverse Pyramid Structure and Dense MLP Block \thanks{A preprint has previously been published in arXiv[39] }}
\titlerunning{PIPMN}
\author{Yunhao Chen\inst{1}\orcidID{0000-0002-8134-2314} \and
Yunjie Zhu\inst{2} \and
Zihui Yan\inst{1},
Yifan Huang\inst{1},
Zhen Ren\inst{1},
Jianlu Shen\inst{1},
Lifang Chen\inst{1}
\thanks{Yunhao Chen, Yunjie Zhu and Zihui Yan contributed equally to this work and should be considered co-first authors.}}
\authorrunning{Y. Chen et al.}
\institute{Jiangnan University, Wuxi, 214000, China \and
University of Leeds, Leeds, LS2 9JT, United Kingdom}
\maketitle              % typeset the header of the contribution
\begin{abstract}

Recently, massive architectures based on Convolutional Neural Network (CNN) and self-attention mechanisms have become necessary for audio classification. While these techniques are state-of-the-art, these works’ effectiveness can only be guaranteed with huge computational costs and parameters, large amounts of data augmentation, transfer from large datasets and some other tricks. By utilizing the lightweight nature of audio, we propose an efficient network structure called Paired Inverse Pyramid Structure (PIP) and a network called Paired Inverse Pyramid Structure MLP Network (PIPMN) to overcome these problems. The PIPMN reaches 95.5\% of Environmental Sound Classification (ESC) accuracy on the UrbanSound8K dataset and 93.2\% of Music Genre Classification (MGC) on the GTAZN dataset, with only 1 million parameters. Both of the results are achieved without data augmentation or transfer learning. The PIPMN can achieve similar or even exceeds other state-of-the-art models with much less parameters under this setting. The Code is available on the \url{https://github.com/JNAIC/PIPMN}

\end{abstract}

\section{Introduction}
\label{sec:intro}

Audio classification aims to categorize sounds into several predefined groups such as Environmental Sound Classification (ESC) [1] or Music Genre Classification (MGC) tasks [2]. The demand for accurate audio classification systems has grown in recent years, with applications in various fields like hearing aids [3], urban planning [4], and biology [5]. For instance, MGC can be utilized in music recommendation systems or emotional tests [6,7].

  Meanwhile, deep learning techniques have revolutionized the field of audio classification, showcasing outstanding accuracy. Specifically, Convolutional Neural Networks (CNN) and self-attention mechanisms have been highly effective. These models typically process 2D spectrograms extracted from audio as if they were images [8-10,37]. Additionally, techniques such as transfer learning [11], knowledge distillation [12], and cross-modal cooperative learning [13,37,38] have been employed to enhance model robustness and accuracy.

    However, treating audio data as images has its drawbacks. Unlike 3D images (with channels), the audio spectrum is 2D data. By treating the audio spectrum as an image, both the frequency or cepstral coefficients domain (referred to as the depth domain) and the temporal domain are processed as a spatial domain. This approach necessitates the addition of a new dimension for the input data as a channel to fit the input data into the image model, increasing the computational cost and propensity for overfitting. 
     Moreover, despite improvements in the ESC task by previous works [14-16], these networks primarily focus on the spatial information of the time and frequency domain, often neglecting adequate extraction of depth domain information, resulting in lower accuracy.

  On a more optimistic note, we have observed that the humble Multi-Layer Perceptron (MLP), a simple and computationally efficient module, has demonstrated competitiveness in image classification tasks [17] on ImageNet [18]. Such performance lends credibility to the MLP's ability to process complex information. Inspired by this, we decided to build our network around the MLP. Yet, we recognized that the multi-stage structure [19] commonly used for CNN does not suit MLP-based networks due to the propensity for overfitting. To tackle this, we introduce a long-range skip connection with layer scale [20] on the bottleneck as a countermeasure against overfitting.

In this paper, we aim to establish a novel approach for depth and time domain information extraction in audio classification, emphasizing minimal parameter use. Our motivation stems from three primary aspects. Firstly, we aim to process audio in a lightweight manner rather than treating it as an image. Secondly, we seek to replace CNN and transformer with MLP to reduce computational cost and parameters. This has led to the development of the Dense MLP (DM) block for audio classification. Lastly, to minimize network overfitting, we propose the Paired Inverse Pyramid Structure (PIP).

  To summarize, the main contributions of this paper are as follows:
\begin{itemize}
  \item [1)]
  We propose  Temporal MLP, Depth Domain Block  and Dense MLP block to extract audio's spatial and temporal domain information more effectively.
  \item [2)]
  We propose a Paired Inverse Pyramid Structure to reduce overfitting for multi-stage MLP networks in audio classification.
  \item [3)]
  The entire PIPMN achieves outstanding accuracy on both the UrbanSound8K dataset for the ESC task and the GTAZN dataset for the MGC task without data augmentation and transfer learning.
\end{itemize}

  Section 2 concentrated on the mathematical description and network structures proposed in this paper. Section 3 describes experiment methods and results compared with other high-performing models. Section 4 is the discussion and summary of all the above experiments. Sections 5 and 6 provide a more detailed explanation on the advantages of our model compared with other SOTA models. Section 7 summarizes all the contents. The rest sections include data availability and references.

\begin{figure}
\centering
\includegraphics[width=1.05\textwidth]{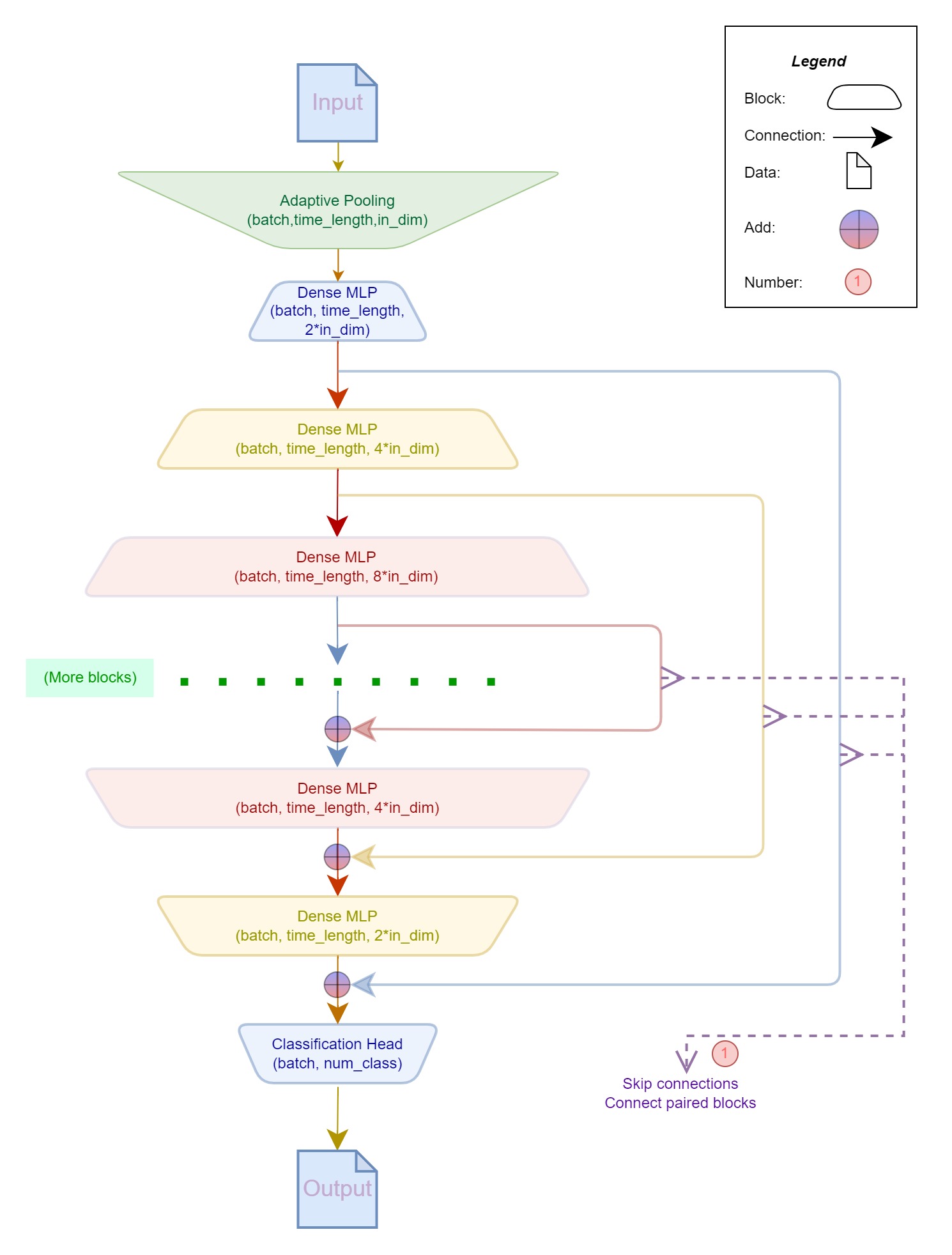}
\caption{Paired Inverse Pyramid Structure.An example of the proposed multi-stage structure with long-range skip connection and layer scale. The structure consists of three pairs of layers with different expansion rates. The long-range skip connection connects the first and the last layer of each pair. The layer scale controls the training dynamics of each layer. The Adaptive Pooling reduces the time domain’s dimension to a fixed length.} \label{fig1}
\end{figure}

\section{METHODS}
\label{sec:format}

\subsection{Paired Inverse Pyramid Structure}

The overall structure of the model is depicted in Figure 1. One of the key features of this structure is the incorporation of a long-range skip connection with layer scale on the bottleneck, which is instrumental in reducing overfitting. The rationale behind this is that as the network becomes deeper in the conventional multi-stage structure, it represents a more complex function due to the increased number of parameters. This complexity enhances the propensity of the network to overfit. However, by utilizing training dynamics [20] offered by layer scale, the network can represent a less complex function from the skip connection, making it less susceptible to overfitting.

Moreover, previous studies on deep networks with stochastic depth [21] demonstrate that the learning of a layer depends on the information obtained from both the preceding layer and the non-adjacent layer. Therefore, the long-range skip connection can preserve the information extracted by a layer and transmit it to another layer that requires it, without losing it in multiple intermediate layers.

The whole structure can be described as follows:

\begin{equation}\label{eq}
    \varpi_n=[\kappa_1,\kappa_2,...,\kappa_{n-1} ,\kappa_n,\kappa_{n-1},...,\kappa _1 ]
\end{equation}

Where $n$ is the number of pairs of layers, $\kappa_i$  is the hyperparameter which is the expansion rate  of the depth domain's dimension. An example of this structure is demonstrated in Fig 1 where $n=3$, $\kappa_1=2,\kappa_2=4,\kappa_3=8$ and $\varpi_3=[2,4,8,4,2]$. The $(Batch,time\_length,\varpi[i] \times in\_dim)$ means the size of the sensor after processing by the layer. The $Batch$ is the batch size of the input tensor, the $time\_length$ is the hyperparameter set for Adaptive Pooling and $in\_dim$ is the initial depth domain's dimension. 

\subsection{Positional Modelling for Time Domain and Depth Domain}
The spectrogram fed into the positional modelling can be represented as $\chi \in \mathbb{R}^{ B\times D \times L}$ where $B$ is the batch size, $L$ is the temporal domain, and $D$ is the Depth domain. And the positional modelling for the depth domain on the time domain can be described as follows:
\begin{equation}\label{eq1}
\varphi(\chi) = \chi * \gamma_{3}^{1}
\end{equation}
where $\varphi(\chi)$ is the positional modelling, the $\gamma_{3}^{1}$ is the convolution layer with a kernel size of 3 and stride of 1. To fit the positional modelling to the network, we used the 1-padding in the layer. To reduce parameters, the group of input equals $D$.

\subsection{ Temporal MLP Based on Feed-Forward Structure}
The structure is illustrated in Figure 2. The feed-forward structure includes one skip connection [22], one MLP block and a positional modelling block. The MLP block includes one LayerNorm [23] layer, two Linear layers and a GELU [24] activation function. 

The operation of the MLP block is akin to the structure of a transformer [25]. In the MLP block, the length of the time domain is first expanded and then condensed back to its original length. This mechanism aids in the optimal exploitation of information within the time domain. The degree of expansion is regulated by a hyperparameter denoted as $\alpha$.

 The MLP block can be described using the following equation:
\begin{equation}\label{eq2}
\phi(\chi)=W_{L\times \alpha L}^{T}(GELU(W_{\alpha L \times L}^{T}(\eta(\chi))+b_{1}))+b_{2}
\end{equation}
where $\chi\in\mathbb{R}^{B\times D \times L} $ ,$\eta(\chi)$   is the layer normalization for $\chi$ , $W_{a\times b}$ is the learnable parameters vector of size $a\times b$, $b_{1}$, $b_{2}$ is the corresponding bias, GELU is the activation function and $L=time\_length$.
The whole feed-forward structure can be described as follows:
\begin{equation}\label{eq3}
\gamma(\chi)=W_{3\alpha L \times L}^{T}([\varphi(\chi),\chi,\phi(\chi)])
\end{equation}
where $\chi\in\mathbb{R}^{B\times D \times L} $,$\phi(\chi)$ is the operation in the MLP, [.,.,.] is the concatenation operation between tensors, and $W_{3\alpha L \times L}$ transforms the size of the concatenation tensor into the original one. This feed-forward structure aims to extract the information related to the temporal domain and the relations between the temporal and depth domain.

\begin{figure}
\centering
\includegraphics[width=1.05\textwidth]{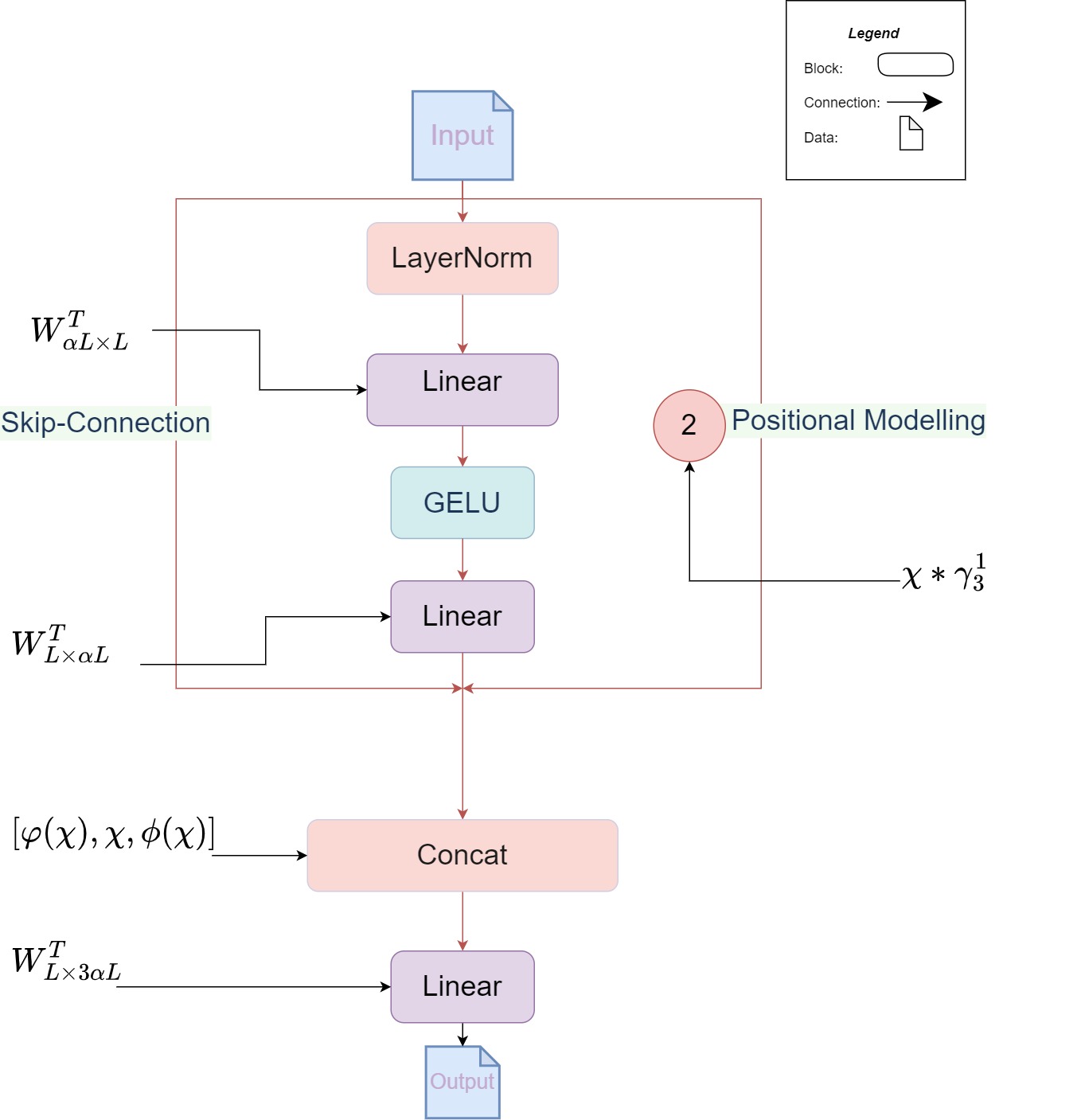}
\caption{The temporal MLP based on feed-forward structure. The structure consists of a skip connection, an MLP block and a positional modelling block. The MLP block expands and shrinks the time domain length using two linear layers and a GELU activation function. The expansion ratio is controlled by a hyperparameter $\alpha$. The positional modelling block concatenates the original input, the MLP output and the skip connection output, and transforms them into the original size using another linear layer. The structure is defined by equations (2), (3) and (4). } \label{fig2}
\end{figure}

\subsection{Depth Domain Block and Linear Skip Connection}
Due to the complicated structure of PIP, the depth domain block of each dense MLP block is relatively simple to reduce overfitting, which will not compromise its overall accuracy and will be less likely to overfit. The structure is depicted in Figure 3 and can be described as follows:
\begin{equation}\label{eq4}
\delta(\chi)=GELU(W_{D_{in}\times D_{out}}^{T}\eta(\chi))+ \omega_{D_{in} \times D_{out}}^{T}(\chi)
\end{equation}
where $w_{D_{in}\times D_{out}}$ is the learnable parameters vector of size $D_{in}\times D_{out}$ and  $\chi\in\mathbb{R}^{B\times L \times D} $ (its size permutation is different from the tensor $\chi$ mentioned above). The special skip connection for this layer deserves attention. Because the output tensor size of this layer differs from the input tensor size, we need to apply a simple Linear layer to the input to match the skip connection with the structure. The skip connection is effective because it provides a simple way to transform and map the original data to the processed data. Without this skip connection, the information propagated by multi-layers will be lost due to multiple transformations.
Moreover, compared with the inherited complex structure of a deep neural network, a simple Linear layer is simple enough to serve as the skip connection in terms of transformation. Therefore, a Linear layer can act as the skip connection implemented with an identity shortcut in the ResNet. We call this module the Linear Skip Connection.

\begin{figure}
\centering
\includegraphics[width=1.2\textwidth]{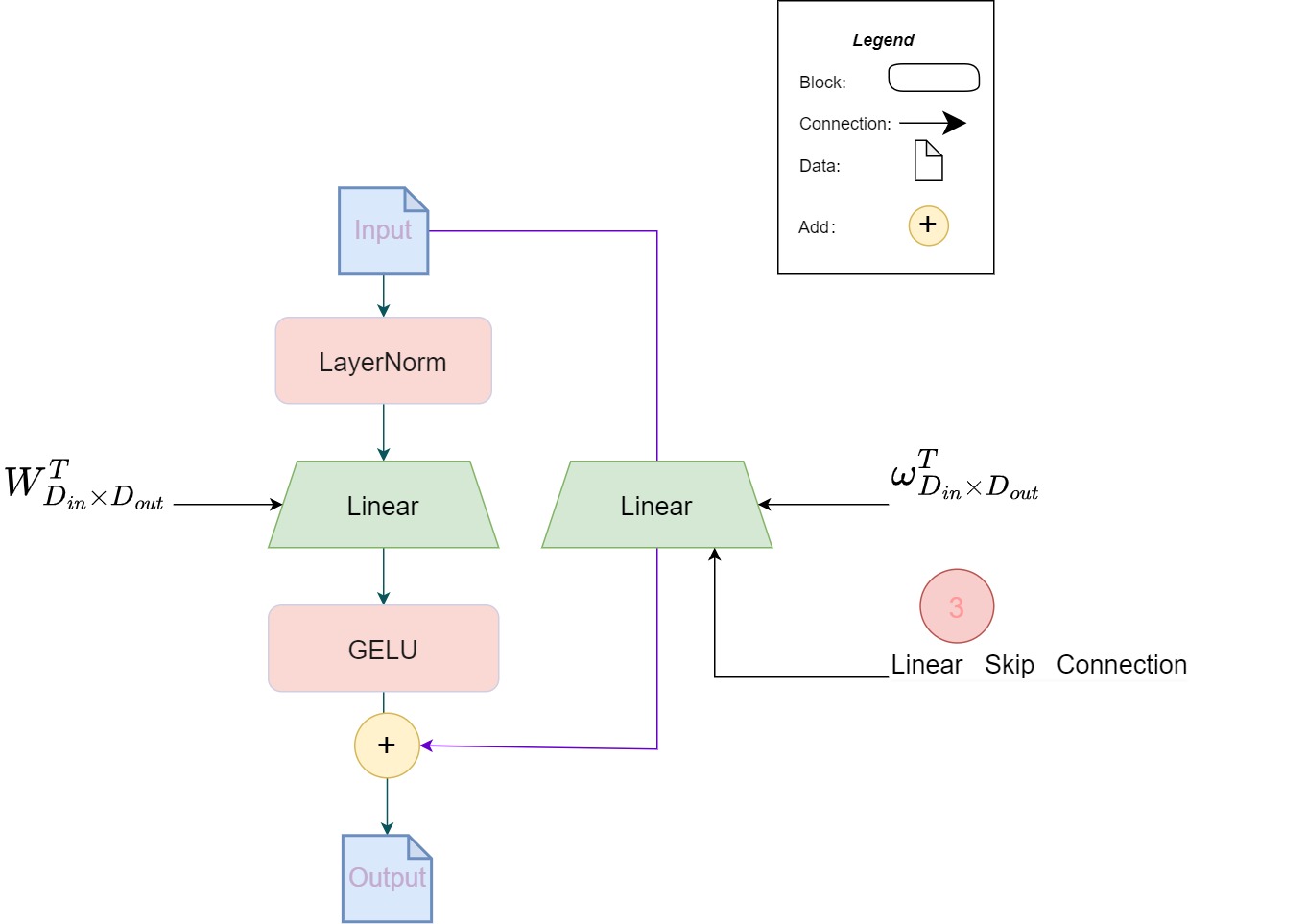}
\caption{The depth domain block of the dense MLP block. The block consists of a layer normalization, a linear layer, a GELU activation function and a linear skip connection. The block transforms the input tensor from size $D_{in}$ to size $D_{out}$ in the depth dimension using equation (5). The linear skip connection adds a linear layer to the input tensor to match the output size and preserve the original information. } \label{fig3}
\end{figure}

\subsection{Dense MLP}
The whole structure is depicted in Figure 4. And the processing procedure for the structure can be described as follows:
\begin{equation}\label{eq5}
\theta(\chi)=\epsilon_{2}\delta(\epsilon_{1} \lambda(\gamma(\lambda(\chi)))+\chi)
\end{equation}
where $\lambda(\chi)$ is the operation of permutation between the depth domain and temporal domain, the operation transforms the $\chi\in\mathbb{R}^{B\times L \times D}$  into  $\chi\in\mathbb{R}^{B\times D \times L}$  or transforms the $\chi\in\mathbb{R}^{B\times D \times L}$into $\chi\in\mathbb{R}^{B\times L \times D}, \epsilon_{1},\epsilon_{2}$ means learnable scalers as the layer scale value[20], which can reduce the overfitting according to the experiments. For example, when $n=2$, $\varpi_1=[2,4,2]$, $time\_length=10$ and $in\_dim=100$, the $D_{in}$ and $D_{out}$ for the first Dense MLP is $in\_dim$ and $2 \times in\_dim$, for the second Dense MLP is $2 \times in\_dim$ and $4 \times in\_dim$, for the third Dense MLP is $4 \times in\_dim$ and $2 \times in\_dim$.

\begin{figure}
\centering
\includegraphics[width=1.0\textwidth]{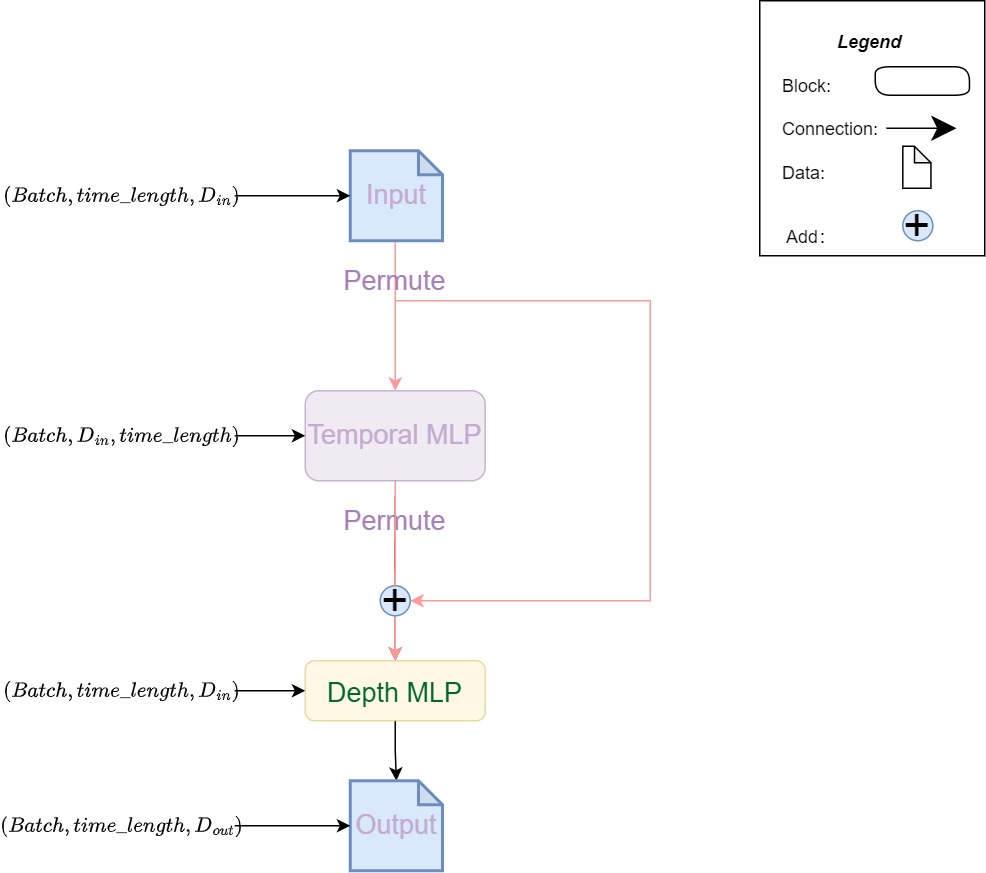}
\caption{The Structure of Dense MLP. The proposed structure for the Dense MLP. The module consists of blocks with different input and output dimensions, followed by a permutation operation between the depth and temporal domains. The structure is defined by equation (6), where $\lambda(\chi)$ is the permutation function and $\epsilon_1$ and $\epsilon_2$ are learnable scalers. } \label{fig4}
\end{figure}

\section{Experiments}
\label{sec:pagestyle}

\subsection{Overview on the UrbanSound8K Dataset}
The UrbanSound8K dataset, created by Salamon et al. in 2014, is an extensive compendium of urban noise recordings. It encompasses a collection of 8732 sound samples, all of which have been meticulously classified into ten distinct categories drawn from a broad urban sound taxonomy. These categories include common urban sounds such as car horns, sirens, and children playing, amongst others.

The samples, gathered from various field recordings, have a maximum duration of 4 seconds and are sampled at varying rates, ranging from 16kHz to 48kHz. The dataset is equipped with additional metadata that provides insightful details like the geographical location where the sample was recorded and the type of device employed for the recording. This dataset has evolved into an essential benchmark for testing and validating the performance of audio classification models, particularly those focused on urban sounds.

\subsection{Overview on the GTZAN Dataset}
The GTZAN dataset, crafted by Tzanetakis and Cook, is a comprehensive dataset dedicated to the domain of music genre classification. Comprising 1000 audio tracks, each 30 seconds long, the dataset presents a balanced representation of ten diverse genres, including blues, classical, country, disco, hip-hop, jazz, metal, pop, reggae, and rock.

All the tracks in the GTZAN dataset adhere to the WAV format, sampled at a frequency of 22.05 kHz and a 16-bit resolution. This extensive and varied compilation of music has been extensively utilized as a benchmark for evaluating the performance of music genre classification models. Besides, it also finds applications in other audio classification tasks, thus proving to be a valuable asset for researchers in the field of music information retrieval (MIR) and audio signal processing.

\subsection{Overview of the FreeSound 2019 Dataset}
The FreeSound 2019 dataset, released in the year 2019, stands as a monumental resource for sound event detection and classification tasks. Hosted on the FreeSound platform, the dataset offers a rich collection of over 41,000 audio recordings, covering a total duration of approximately 500 hours. The recordings are diverse, encompassing 80 categories of sounds, including but not limited to musical instruments, human sounds, animal sounds, and environmental sounds. To better evaluate the performance of the models, we only use the curated training set, which contains 4970 recordings. The curated dataset represents high-quality, carefully selected samples that are representative of the overall data distribution. It provides a consistent, fair, and reliable ground for comparative model evaluation. Consequently, using the curated dataset for comparison adheres to the principle of comparability, which is crucial in any model evaluation.

Each recording in the dataset is sampled at a frequency of 44.1 kHz, and it comes with accompanying annotations that signify the presence of different sound events. These annotations are provided as distinct labels corresponding to each sound event. The FreeSound 2019 dataset is the primary dataset for challenges such as the DCASE 2019 Challenge,  underlining its significance in the research community. 

The Dataset is available on the \url{https://www.kaggle.com/competitions/freesound-audio-tagging-2019/overview}.

\begin{figure}
\centering
\includegraphics[width=0.6\textwidth]{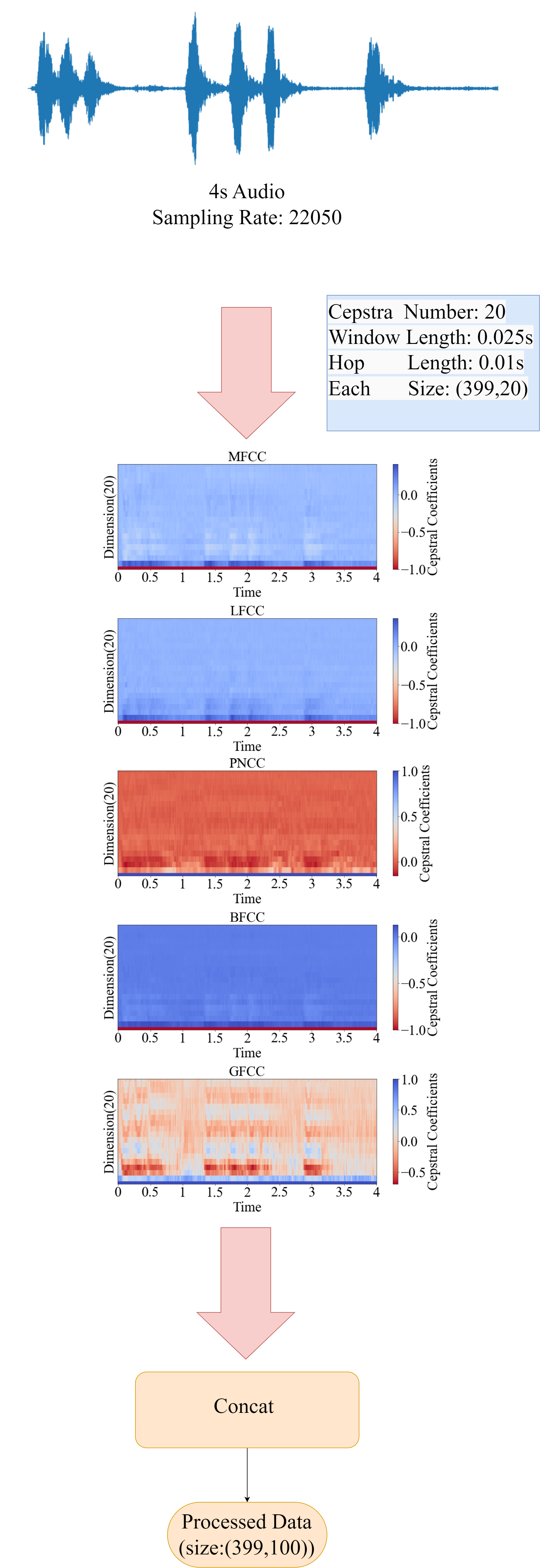}
\caption{Different types of acoustic features are extracted from the audio samples and concatenated into a single tensor. The audio samples are segmented into 4s clips to fit the input size of the model. The input tensor has a shape of (128,399,100), where 128 is the batch size, 399 is the number of frames, and 100 is the feature dimension.} \label{fig5}
\end{figure}

\subsection{Training setup and Preprocessing for the Datasets}

In our study, we evaluated our approach using three different datasets: UrbanSound8K for the ESC, GTZAN for the MGC, and the curated data in the Freesound 2019 dataset. For feature extraction, we used similar techniques for all datasets, including NGCC, MFCC, GFCC, LFCC, and BFCC, with a size of (399,20) for each feature, using a window length of 0.025 seconds and a hop length of 0.01 seconds.  For GTZAN, we divided each sample into seven segments to fit the input size to our model, resulting in audio samples that are four seconds in length for both GTZAN and UrbanSound8K datasets. Our training batch size was 128, resulting in an input size of (128,399,100) for GTZAN, UrbanSound8K and Freesound 2019, as shown in Figure \ref{fig5}. We used the AdamW optimizer with a learning rate of 0.001 and a weight decay of 0.05, with other parameters set to default. The loss function we used was the Cross-entropy-loss with label-smoothing of 0.1. For training, we used 3500 epochs for all datasets, and the training stopped when the training accuracy reached 100\% and the training loss stopped decreasing. To assess the efficiency of our network, we didn't use any data augmentation, transfer learning, EMA, pretraining, or other techniques. We randomly split the datasets into three parts, with 10\% of the samples reserved for testing and 10\% for validation. We used the curated data available in the Freesound 2019 dataset to ensure a fair comparison of the performance of our model with prior studies.

The hyperparameters of PIPMN are set as follows: \\$time\_length=5$, $in\_dim=100$, $n=2$, $\varphi_2=[4,8,4]$ and $\alpha=3$.

The comparison models are implemented by virtual of timm\cite{timm} and their hyperparameters are the default parameters in timm.

\subsection{UrbanSound8K Experiments Results}
% \begin{table}
% \centering
% \caption{PREVIOUS MODELS VS. THE PROPOSED MODEL IN THIS PAPER ON URBANSOUND8K DATASET.}\label{tab1}
% \begin{tabular}{|l|l|l|}
% \hline
% Framework& Classification Accuracy(\%) &  Parameters & Ref.\\
% \hline
%    VIT-Small & 93.7 & 21.5M & [30]\\
%    CoaT-lite-mini &94.4   &10.5M & [31]\\
%    ConViT-Small  &95.0  &27.1M& [32]\\
%    MobileViT III &  95.1    & 21.5M & [33]\\
%    CovNeXT & 94.7 & 87.6M & [19]\\ 
%    Mlp-Mixer & 94.5 & 17.5M & [17]\\
%    Proposed model & 95.5 & 1.4M & \\
% \hline
% \end{tabular}
% \end{table}

% \begin{table}[h]
% \caption{PREVIOUS MODELS VS. THE PROPOSED MODEL IN THIS PAPER ON URBANSOUND8K DATASET}
% \label{table1}

% % \setlength{\tabcolsep}{0pt}
% \begin{tabular}{cp{5cm}{}cp{5cm}{}cp{5cm}{}}
% \hline
% Framework& 
% Classification Accuracy(\%)&Parameters&Ref. \\
% \hline

%    VIT-Small & 93.7 & 21.5M & [30]\\
%    CoaT-lite-mini &94.4   &10.5M & [31]\\
%    ConViT-Small  &95.0  &27.1M& [32]\\
%    MobileViT III &  95.1    & 21.5M & [33]\\
%    CovNeXT & 94.7 & 87.6M & [19]\\ 
%    Mlp-Mixer & 94.5 & 17.5M & [17]\\
%    Proposed model & 95.5 & 1.4M & -\\
% \hline

% \end{tabular}
% \label{tab1}
% \end{table}
% \usepackage{tabularray}
\begin{table}
\centering
\caption{PREVIOUS MODELS VS. THE PROPOSED MODEL IN THIS PAPER ON URBANSOUND8K DATASET. (The unit of measurement for this table is \%. MaP stands for Macro Precision, MaF1 stands for Macro F1, \textbf{LSTM here stands for "Replace DenseMLP with LSTM"}, and MiF1 stands for Micro F1.)}
\label{table1}

\begin{tblr}{
  width = \linewidth,
  colspec = {Q[267]Q[144]Q[154]Q[110]Q[129]Q[123]},
  cell{1}{1} = {c},
  cell{2}{1} = {c},
  cell{2}{3} = {c},
  cell{2}{4} = {c},
  cell{3}{1} = {c},
  cell{3}{3} = {c},
  cell{3}{4} = {c},
  cell{4}{1} = {c},
  cell{4}{3} = {c},
  cell{4}{4} = {c},
  cell{5}{1} = {c},
  cell{5}{3} = {c},
  cell{5}{4} = {c},
  cell{6}{1} = {c},
  cell{6}{3} = {c},
  cell{6}{4} = {c},
  cell{7}{1} = {c},
  cell{7}{3} = {c},
  cell{7}{4} = {c},
  cell{11}{1} = {c},
  cell{11}{3} = {c},
  cell{11}{4} = {c},
  hline{1-2,12} = {-}{},
}
Framework          & Accuracy           & ~ ~ MaP           & MaF1          & ~ ~ ~MiF1                & Params                 \\
VIT-Small[30]      & ~ ~ 93.7           & 93.5          & 93.1          & ~ ~ ~ 91.3          & 21.5M                  \\
CoaT-lite-mini[31] & ~ ~ 94.4           & 94.3          & 94.6          & ~ ~ ~ 92.6          & 10.5M                  \\
ConViT-Small[32]   & ~ ~ 95.0           & 94.5          & 94.5          & ~ ~ ~ 93.9          & 27.1M                  \\
MobileViT III[33]  & ~ ~ 95.1           & 95.3          & 95.2          & ~ ~ ~ 95.0          & 21.5M                  \\
CovNeXT[19]        & ~ ~ 94.7           & 95.4          & 95.1          & ~ ~ ~ 93.6          & 87.6M                  \\
Mlp-Mixer[17]      & ~ ~ 94.5           & 94.6              & 94.9                & ~ ~ ~ 94.3                    & 17.5M                  \\
~  ~ LSTM[42]    & ~ ~ 91.6           & ~ ~ ~92.2 & ~ 91.6      & ~ ~ ~ 90.6          & 11.0M                  \\
~ ~CovNeXTV2[43]   &~ ~ 94.9                    & ~ ~ ~95.7             & ~ 95.4               &~ ~ ~ 94.6                     & 49.5M                  \\
~ ~AST[44] & ~ ~ 94.7           & ~ ~ ~95.5 & ~ 95.1      & ~ ~ ~ \textbf{95.3} & 85.5M                  \\
Proposed model     & ~ ~ \textbf{95.5 } & \textbf{95.8} & \textbf{95.9} & ~ ~ ~ 95.1          & \textbf{\textbf{1.4M}} 
\end{tblr}

\end{table}

Table 1 compares our model's effectiveness with other high-performing models on the UrbanSound8K dataset. The table shows that our proposed model achieves higher results with fewer parameters. Consequently, our model can retrieve helpful information without too much overfitting.

Figure 5 and 6  separately present the loss and accuracy curve diagram.  The curve diagrams of accuracy and loss of the training set and validation set demonstrate the model is easy to train because the curve is smooth without sudden changes. 

\begin{figure}
\centering
\includegraphics[width=\textwidth]{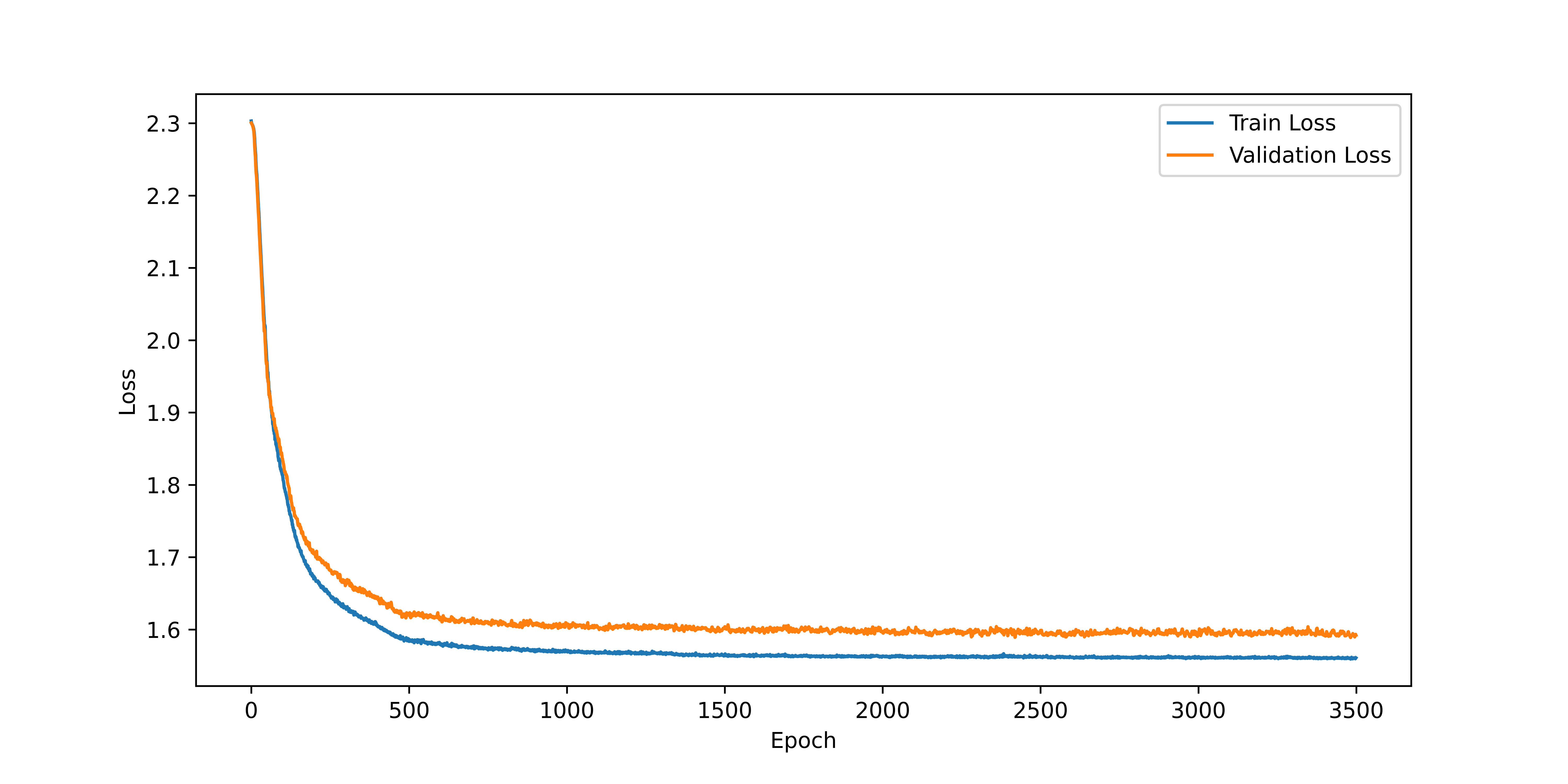}
\caption{The loss curve diagram of the training set and validation set. The smooth curve indicates that the model is easy to train without sudden changes in loss.} \label{fig6}
\end{figure}
\begin{figure}
\centering
\includegraphics[width=\textwidth]{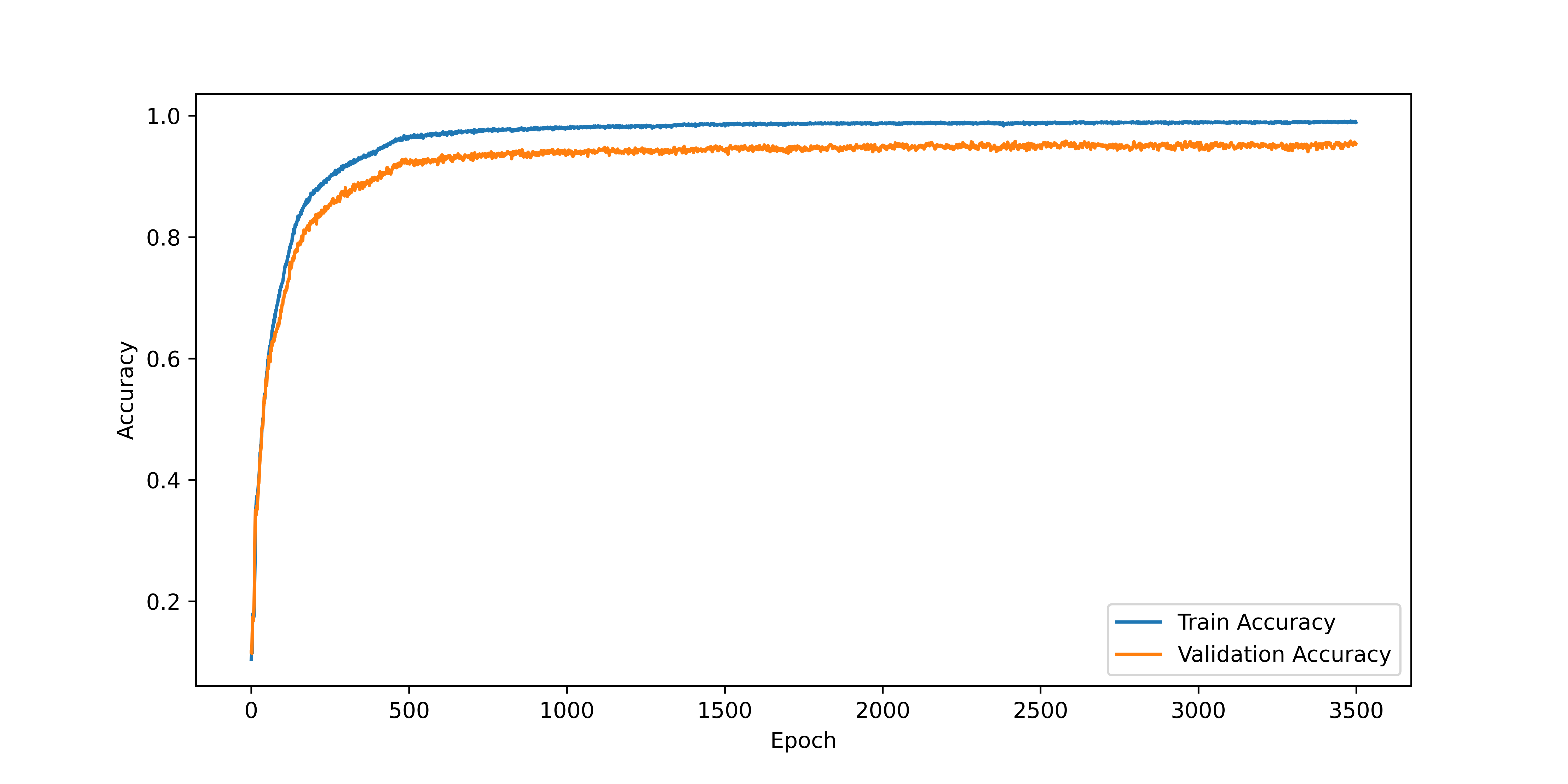}
\caption{The accuracy curve diagram of the training set and validation set. The smooth curve indicates that the model is easy to train without sudden changes in accuracy.} \label{fig7}
\end{figure}

\subsection{GTZAN Experiments Results}
To further explain the efficiency of our model, we evaluate our model on the GTZAN dataset for the MGC task.\\
% \begin{table}
% \centering
% \caption{PREVIOUS STATE-OF-THE-ART ESC MODELS VS. THE PROPOSED MODEL IN
% THIS PAPER ON GTAZN DATASET}\label{tab2}
% \begin{tabular}{|l|l|l|}
% \hline
% Framework & Classification Accuracy(\%) &  Parameters & Ref.\\
% \hline
%    VGGish & 92.2 & 72.1M & [34]\\
%    VGGish+CoTrans-b2 &95&72.6M & [34]\\
%    Improved-BBNN & 91 & 0.18M & [35-36]\\
%    Proposed model & 93.3 & 1.4M & -\\
% \hline
% \end{tabular}
% \end{table}
% \usepackage{tabularray}

\begin{table}
\centering
\caption{PREVIOUS STATE-OF-THE-ART ESC MODELS VS. THE PROPOSED MODEL INTHIS PAPER ON GTAZN DATASET(The unit of measurement for this table is %. MaP
stands for Macro Precision, MaF1 stands for Macro F1, \textbf{LSTM here stands for "Replace DenseMLP with LSTM"}, and MiF1 stands for Micro
F1.)}
\label{table2}
\begin{tblr}{
  width = \linewidth,
  colspec = {Q[377]Q[150]Q[83]Q[98]Q[88]Q[129]},
  column{1} = {c},
  column{6} = {c},
  hline{1-2,6} = {-}{},
}
Framework             & Accuracy        & MaP           & MaF1          & MiF1          & Params          \\
VGGish[34]            & ~ ~ 92.2        & 93.2          & 92.8          & 93.0          & 72.1M           \\
VGGish+CoTrans-b2[34] & ~ ~~\textbf{95} & \textbf{95.6} & \textbf{93.2} & \textbf{94.7} & 72.6M           \\
Improved-BBNN[35-36]  & ~ ~ 91          & 91.2          & 91.4          & 91.5          & \textbf{0.18M } \\
Proposed model        & ~ ~93.3         & 93.0          & 93.4          & 93.7          & 1.4M            
\end{tblr}
% \begin{tblr}{
%   width = \linewidth,
%   colspec = {Q[292]Q[367]Q[167]Q[106]},
%   column{odd} = {c},
%   hline{1-2,6} = {-}{},
% }
% Framework         & Classification Accuracy(\%) & Parameters & Ref.      \\
% VGGish            & ~ ~ ~ ~ ~ ~ ~ ~ ~92.2       & 72.1M      & {[}34]    \\
% VGGish+CoTrans-b2 & ~ ~ ~ ~ ~ ~ ~ ~ ~\textbf{95}         & 72.6M      & {[}34]    \\
% Improved-BBNN     & ~ ~ ~ ~ ~ ~ ~ ~ ~91         & \textbf{0.18M }     & {[}35-36] \\
% Proposed model    & ~ ~ ~ ~ ~ ~ ~ ~ ~93.3       & 1.4M       & -         
% \end{tblr}
\end{table}

As we can see from Table 2, though our model is not state-of-art or the most lightweight model, our model maintains a balance between the parameters and the accuracy. What is more, these models are specialized for the MGC task. Compared with other models, our model is good at both the ESC and MGC tasks. Consequently, we can confidently conclude that our model has scalability in the audio classification.  

\subsection{FreeSound 2019 Experiments Results}
The presented experiment in Table 3 showcases the performance comparison of various models on the FreeSound2019 dataset. The performance is evaluated using three metrics: Example-based Accuracy (EA), Label-based Macro Accuracy (LMaA) and Label-based Micro F1 score (LMiF1). The model size is also considered, indicated by the number of parameters (Params).

In summary, while some models achieved higher LMaA and LMiF1, the proposed model demonstrated the highest EA scores. Moreover, the proposed model achieved these results with significantly fewer parameters, implying superior efficiency and lower computational requirements, which is a substantial advantage in practical applications.

\begin{table}
\centering
\label{table4}
\caption{PREVIOUS MODELS VS. THE PROPOSED MODEL IN THIS PAPER
ON FREESOUND2019 DATASET. (The unit of measurement for this table is \%. EA stands for Example-based Accuracy, LMaA stands for  Label-based Macro Accuracy, LMiF1 stands for Label-based Micro F1 score, 
 and \textbf{LSTM here stands for "Replace DenseMLP with LSTM}",)}
\centering
\begin{tblr}{
  width = \linewidth,
  colspec = {Q[340]Q[133]Q[125]Q[173]Q[146]},
  cell{1}{1} = {c},
  cell{2}{1} = {c},
  cell{2}{3} = {c},
  cell{3}{1} = {c},
  cell{3}{3} = {c},
  cell{4}{1} = {c},
  cell{4}{3} = {c},
  cell{7}{1} = {c},
  cell{7}{3} = {c},
  cell{10}{1} = {c},
  cell{10}{3} = {c},
  hline{1-2,11} = {-}{},
}
Framework              & ~ ~ EA            & ~LMaA            & ~ ~ ~LMiF1         & Params                 \\
VIT-Small[30]          & ~ ~ 38.9          & 33.6          & ~ ~ ~39.3          & 21.5M                  \\
CoaT-lite-mini[31]     & ~ ~ 23.7          & 19.7           & ~ ~ ~24.2          & 10.5M                  \\
ConViT-Small[32]       & ~ ~ 34.7          & 31.1           & ~ ~ ~35.8          & 27.1M                  \\
~ ~ ~MobileViT III[33] & ~ ~ 28.7                 & ~ ~ 22.1                & ~ ~ ~30.0                    & 21.5M                  \\
~ ~ ~CovNeXTV2[43]     & ~ ~ 43.1          & ~ ~~\textbf{36.0} & ~ ~ ~\textbf{45.4} & 87.6M                  \\
Mlp-Mixer[17]          & ~ ~ 37.1          & ~28.6          & ~ ~ ~38.3          & 17.5M                  \\
~ ~ ~ ~ LSTM[42]         & ~ ~ 40.5          & ~ ~ 27.6          & ~ ~ ~37.6          & 11.0M                  \\
~ ~ ~ ~ AST[44]    & ~ ~ 39.1          & ~ ~ 34.2          & ~ ~ ~40.0          & 85.5M                  \\
Proposed model         & ~ ~ \textbf{43.2} & ~31.0          & ~ ~ ~41.2 & \textbf{\textbf{1.4M}} 
\end{tblr}
\end{table}

\subsection{Abalation Study}
\begin{table}
\centering
\caption{ABLATION STUDY ON URBANSOUND8K DATASET(The unit of measurement for this table is \%. MaP stands for Macro Precision, MaF1 stands for Macro F1, MiF1 stands for Micro F1, DM stands for DenseMLP,  PIPS stands for Paired Inverse Pyramid Structure, and OMS stands for Original Multi-Stage.)}
\label{table3}
\begin{tblr}{
  width = \linewidth,
  colspec = {Q[350]Q[160]Q[106]Q[106]Q[100]Q[113]},
  cell{1}{1} = {c},
  cell{1}{6} = {c},
  cell{2}{1} = {c},
  cell{2}{6} = {c},
  cell{3}{1} = {c},
  cell{3}{6} = {c},
  cell{4}{1} = {c},
  cell{4}{6} = {c},
  cell{5}{1} = {c},
  cell{5}{6} = {c},
  cell{6}{1} = {c},
  cell{6}{6} = {c},
  cell{7}{1} = {c},
  cell{7}{6} = {c},
  cell{9}{1} = {c},
  cell{9}{6} = {c},
  cell{10}{1} = {c=6}{0.934\linewidth},
  hline{1-2,10} = {-}{},
}
Framework                                                                                                                                                                  & ~ ~ ~ ~ ~ ~Accuracy         & ~ ~ ~ ~Map          & ~ ~ ~ MaF1           & ~ ~ ~ MiF1           & Params       \\
Without \textcircled{1}                                                                                                                                                                  & ~ ~ ~ ~ ~ ~ ~ 95.1          & ~ ~ ~ 95.2          & ~ ~ ~ ~95.2          & ~ ~ ~ ~94.6          & 1.4M         \\
Without \textcircled{2}                                                                                                                                                                  & ~ ~ ~ ~ ~ ~ ~ 95.2          & ~ ~ ~ 95.6          & ~ ~ ~ ~95.5          & ~ ~ ~ ~95.0          & 1.4M         \\
Without \textcircled{3}                                                                                                                                                                  & ~ ~ ~ ~ ~ ~ ~ 93.6          & ~ ~ ~ 93.7          & ~ ~ ~ ~94.0          & ~ ~ ~ ~93.2          & 1.4M         \\
MFCC                                                                                                                                                                       & ~ ~ ~ ~ ~ ~ ~ 95.2          & ~ ~ ~ 95.4          & ~ ~ ~ ~95.5          & ~ ~ ~ ~95.1          & 0.3M         \\
Mel-Spectrogram                                                                                                                                                            & ~ ~ ~ ~ ~ ~ ~ 92.1          & ~ ~ ~ 92.5          & ~ ~ ~ ~92.7          & ~ ~ ~ ~91.9          & 1.4M         \\
Replace PIPS with OMS                                                                                                                                     & ~ ~ ~ ~ ~ ~ ~ 93.9          & ~ ~ ~ 93.7          & ~ ~ ~ ~94.1          & ~ ~ ~ ~94.2          & 2.4M         \\
~  Replace DM with LSTM                                                                                                                                        & ~ ~ ~ ~ ~ ~ ~ 91.6~         & ~ ~ ~ 92.2~         & ~ ~ ~ ~91.6~         & ~ ~ ~ ~90.6          & ~ 11.0M \\
Proposed model                                                                                                                                                             & ~ ~ ~ ~ ~ ~ ~ \textbf{95.5} & ~ ~ ~ \textbf{95.8} & ~ ~ ~ \textbf{~95.9} & ~ ~ ~ ~\textbf{95.1} & 1.4M         \\
\textcircled{1} is Long Range Skip Connection, \textcircled{2} is Positional Modelling, \textcircled{3} is Linear Skip Connection. The structures represented by these numbers are already indicated in the figures. &                             &                     &                      &                      &              
\end{tblr}

\end{table}

The table \ref{table3} presented showcases an ablation study on the UrbanSound8K dataset to delve into the influence of various components on the overall classification results. Two experiments were conducted to examine the effects of different data preprocessing techniques. The first experiment involved replacing the original data input with 50 Mel Frequency Cepstral Coefficients (MFCCs). The second experiment entailed replacing the input with a 100-Mel-Spectrogram. The results from these experiments suggest that altering the input does not significantly impact the performance of our model, implying that our model's results are not strongly tied to the specific type of input used.

Moreover, the experiment "Replace DM with LSTM" refers to a test where the Dense MLP (DM) block in the model was substituted with a Long Short-Term Memory (LSTM) unit. LSTMs, a type of recurrent neural network, are adept at learning long-term dependencies in data, especially in time-series data like audio. However, they can be computationally demanding and require a larger number of parameters, potentially leading to longer training times and overfitting issues. In this experiment, the substitution aimed to test the relative effectiveness of LSTM against the Dense MLP block in the model. However, the results indicated a reduction in performance metrics such as accuracy, MAP, MaF1, and MiF1 when LSTM was used, in addition to an increase in the model's complexity due to a larger number of parameters (from 1.4M to 11.0M). In summary, replacing the Dense MLP with LSTM resulted in lower performance and increased complexity, indicating that the Dense MLP block in the PIPMN model is more effective and efficient for this particular task and dataset.

Furthermore, to assess the effectiveness of the Paired Inverse Pyramid Structure (PIPS), we replaced the PIPS in our model with a conventional multi-stage structure. Additionally, we conducted an evaluation of the distinct modules in the Dense Multi-Layer Perceptron (MLP) by sequentially removing elements \textcircled{1}, \textcircled{2}, and \textcircled{3}, where \textcircled{1} denotes Long Range Skip Connection, \textcircled{2} represents Positional Modelling, and \textcircled{3} signifies Linear Skip Connection. The outcomes of these experimental modifications are displayed in Table 3. The results affirm the effectiveness and integral role of these three components in our PIPMN model.

The original multi-stage structure is visualized in Figure 8, which excludes the transition layer for simplicity. It is apparent from a theoretical perspective that the original multi-stage structure contains a significantly larger number of parameters compared to the PIPS. Table III demonstrates the better performance of our PIPS when compared with the original multi-stage structure. Our proposed PIPMN model consistently outperforms the original structure, asserting its efficacy in audio classification tasks.

\begin{figure}
\centering
\includegraphics[width=1.0\textwidth]{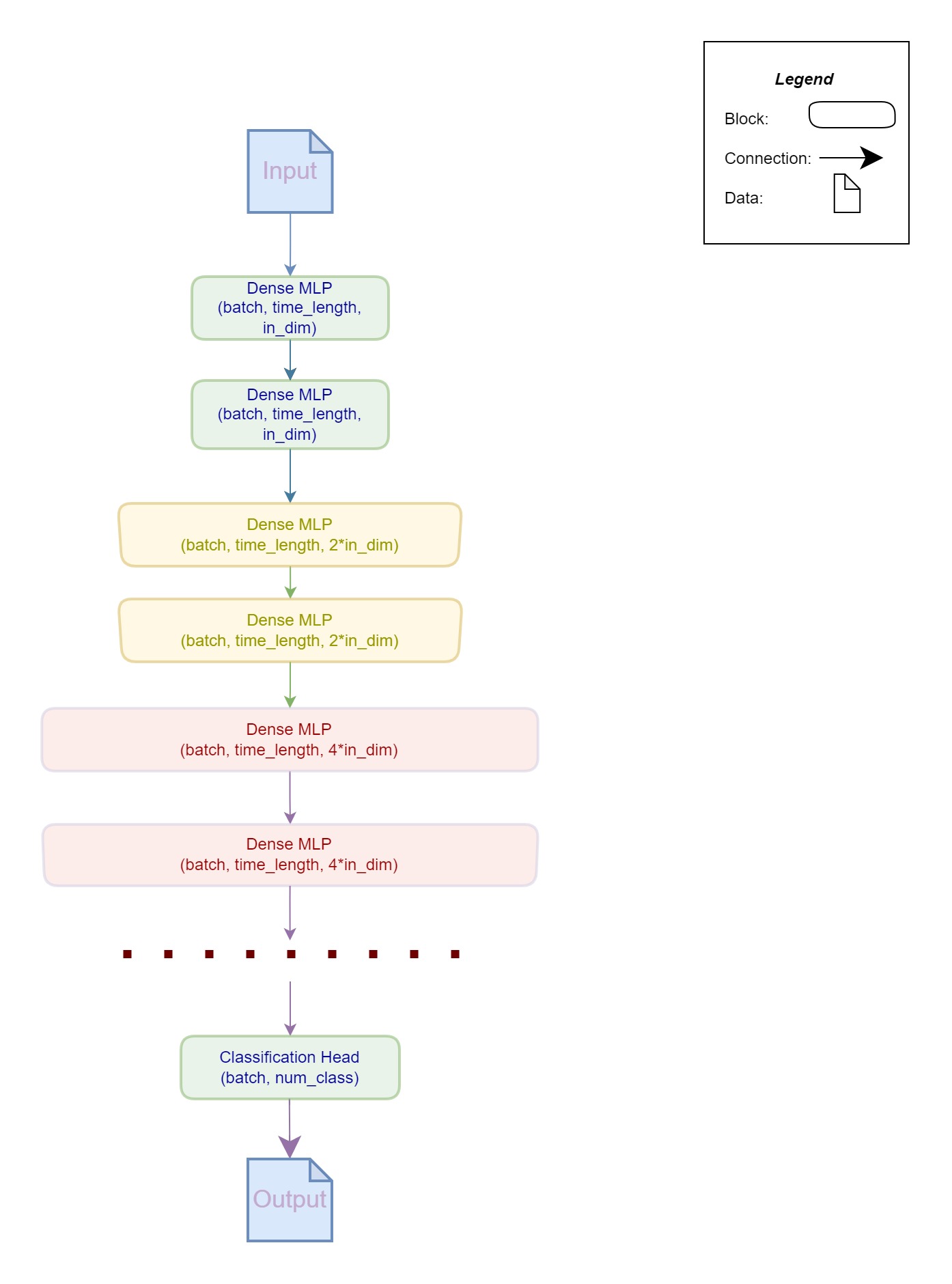}
\caption{Multi-Stage Structure as Comparison Structure to Demonstrate the Effectiveness of the PIP Structure} \label{fig8}
\end{figure}

\section{Discussion}
In this paper, we applied our model to the UrbanSound8K dataset symbolizing environmental sound classification and to the GTAZN dataset representing music genre classification. The experiments verified that the PIPMN has better generalization abilities than high-performing classification models that process audio spectrogram when data augmentation and transfer learning  are unavailable. 

We have shown that the long-range skip connection with layer scale on the bottleneck(PIP Structure) can reduce overfitting and improve performance based on the ablation study. The reasons behind this effect are from the perspective of function complexity and information preservation. The long-range skip connection allows the network to fit from a less complex function represented by the skip connection, which reduces the risk of overfitting. The layer scale controls the training dynamics of each layer and adjusts the contribution of the skip connection and the original structure. The information extracted by a layer can be transmitted to another that needs it without losing it in multiple intermediate layers. This enhances the information flow and the learning ability of the network. Our experiments on different datasets and tasks have validated our claims and demonstrated the effectiveness and robustness of our structure. We believe that our structure can be applied to other deep neural networks and benefit various applications that require high accuracy and low complexity.

We have also demonstrated the effectiveness of the Linear Skip Connection by virtual of experiments. The reasons are as follows: firstly, the Linear Skip Connection applies a simple Linear layer to the input to match the output tensor size of the layer. This enables a simple way to transform and map the original data to the processed data. Without this Linear Skip Connection, the network's information flow and learning ability would be impaired by multiple transformations. Furthermore, compared with the inherited complex structure of a deep neural network, a simple Linear layer is simple enough to function as the skip connection in terms of transformation. Therefore, a Linear layer can emulate the skip connection implemented with an identity shortcut in the ResNet. The results of our experiments on various datasets and tasks have confirmed our arguments and shown the efficiency and reliability of our Linear Skip Connection. We propose extending our Linear Skip Connection to other deep neural networks and enhancing various applications that demand high accuracy and low complexity.

While it is true that MLPs are fully connected and may result in a higher number of parameters compared to CNNs or Transformers, the overall model complexity in this work remains lower. The PIPMN can have a much smaller number of layers without sacrificing good performance, resulting in lower computational cost and memory requirements. This is because the input data has already been preprocessed and engineered with cepstral coefficients. Consequently, an MLP-based structure is sufficient for the task, as verified in the ablation study and experiments. When the input data is replaced with the Mel-Spectrogram, the accuracy of PIPMN decreases dramatically. However, PIPMN can achieve better accuracy than other state-of-the-art models when the input data consists of cepstral coefficients. Consequently, the preprocessed input data allows for simpler MLP-based structures with fewer layers, and the PIPMN is sufficient for the preprocessed and engineered input data.

In addition, although CNNs have the advantage of translation invariance and weight sharing, which allows them to learn more robust features, they are primarily designed for spatial data, such as images. However, not all data exhibit strong spatial patterns. In this work, where the audio is transformed into cepstral coefficients, using an MLP-based structure is more appropriate and efficient, as the benefits of using CNNs are not as pronounced due to the absence of strong spatial patterns in the preprocessed audio.

Also, in the current state-of-the-art models, input data is typically treated as three-dimensional data (for instance, in the case of images, which consist of width, height, and channel dimensions). By directly applying such processing to audio spectrograms, which inherently are two-dimensional data, an additional dimension is introduced. This extra dimension inadvertently increases the model's complexity, computational time, and propensity for overfitting. This is verified in the experiments where PIPMN outperforms other state-of-the-art models, which process the input data as three-dimensional, with significantly fewer parameters.

It is also worth noting that our model can exceed those models with much fewer parameters, demonstrating that the proposed model's training speed and inference speed are increased dramatically compared with those high-performing models.  Currently, our research has not compared the PIPMN with models when the data augmentation is available; it is considered to adjust the PIPMN structure or other hyper-parameters in the future.

\section{Why the PIPMN is Better?}
The proposed method in our study, which incorporates the long-range skip connection with layer scale in a bottleneck structure (referred to as PIP Structure), and the use of Linear Skip Connection, offers several advantages over other advanced methods:

\subsection{Efficient Dimension Handling} The Advanced models often treat audio data as 2D images, adding a third dimension for compatibility with image-focused models. This approach increases computational complexity and the potential for overfitting. In contrast, our method respects the inherently two-dimensional nature of audio spectrogram or cepstral coefficients, reducing computational costs and overfitting risks. The two-dimensional nature of audio spectrograms or cepstral coefficients is a key characteristic of audio signals that should be considered in designing an efficient and effective model for audio classification tasks. By respecting this nature, our method can reduce the overfitting risks in several ways:
\subsubsection{Fewer Parameters} By taking into account the two-dimensional nature of the input data, our method can reduce the number of parameters required to process the data. This is because we can use a simpler architecture that operates efficiently on two-dimensional data, which results in a smaller number of parameters. Consequently, we can avoid the need for complex three-dimensional operations that can lead to overfitting due to the increased model complexity.
\subsubsection{Better Modelling of Audio Data} By explicitly modelling the two-dimensional structure of the input data, our method can better capture essential features of the spectrogram or cepstral coefficients that are relevant for audio classification tasks. This can lead to improved accuracy compared to other methods that do not explicitly consider the two-dimensional nature of the data.

Overall, by respecting the inherently two-dimensional nature of audio spectrograms or cepstral coefficients, our method can reduce computational costs, overfitting risks, and improve the accuracy of the model. This is an essential consideration in designing efficient and effective models for audio classification tasks.

\subsection{Effective Information Extraction} While many models focus mainly on the spatial information of the time and frequency domain, our model also efficiently extracts depth domain information, which could lead to improved accuracy. Our Temporal MLP and Depth Domain Block is designed to capture spatial and temporal domain information better. These can be explained from two perspectives:

\subsubsection{Data Representation Perspective} In audio processing, data is often represented in the time-frequency domain, where different frequencies vary over time. Many models primarily capture the spatial relationships within this domain. However, our model goes further to efficiently capture depth domain information. This depth domain might refer to additional features or representations of audio data that aren't just limited to the immediate time-frequency relationships, such as cepstral coefficients, which could provide a richer understanding of the audio signal, leading to improved accuracy. Our Temporal MLP and Depth Domain Block are specifically designed to extract better and utilize this multi-dimensional information.

\subsubsection{Model Architecture Perspective} Many current models leverage architectures designed to focus on the spatial relationships within the time-frequency domain of audio data. They may use convolutional layers or attention mechanisms to understand how different frequencies interact over time. However, our model, with the Temporal MLP and Depth Domain Block, introduces a new approach to architecture design for audio data. These structures are created with a focus on extracting not only the typical time-frequency information but also the depth domain information, which is often overlooked. This unique architectural approach helps our model improve accuracy by comprehensively learning from the audio data.

Overall, the proposed model offers a unique and innovative approach to audio classification that effectively extract information from audio data. This leads to improved accuracy compared to models focusing solely on time-frequency information.

\subsection{Computational Efficiency} We use Multi-Layer Perceptrons (MLPs) instead of more complex structures like Convolutional Neural Networks (CNNs) and transformers. MLPs have demonstrated their effectiveness in tasks such as image classification, and we believe they can similarly excel in audio classification tasks. Using MLPs in our structure allows for reduced computational costs and fewer parameters. This can be explained in several perspectives: 

\subsubsection{Computational Perspective} The use of MLPs in our model, as opposed to more complex structures like CNNs or transformers, brings about a significant reduction in computational costs. CNNs and transformers, due to their complex architectures, require more computing power and resources to train and operate. By utilizing MLPs, which are relatively simpler, we can design models that are easier to train, require less computational power, and yet are capable of achieving high accuracy in tasks like image classification. We believe this efficiency can be similarly applied to audio classification tasks.

\subsubsection{Model Complexity Perspective} MLPs, while simpler than CNNs or transformers, have been shown to be effective in tasks such as image classification. This suggests that a model does not necessarily have to be complex to perform well. Our model leverages the simplicity of MLPs to reduce the total number of parameters, which also leads to a decrease in the likelihood of overfitting. This simpler model can be more interpretable and easier to debug, adjust, and improve.

\section{Why PIPMN is Better than RNNs?}

We appreciate your question regarding the use of recurrent neural networks in processing time-series data. While RNNs are well-suited for time-series data, our proposed method utilizes Multi-Layer Perceptrons (MLPs) instead. We have included a detailed explanation regarding our decision to use MLPs and how they effectively capture the essential features of audio signals.

\subsection{Disadvantages of RNN}

Recurrent Neural Networks (RNNs) have been the traditional choice for handling time-series data due to their inherent capability to model sequential information. However, they come with a set of challenges:

\subsubsection{Vanishing and Exploding Gradients}

RNNs suffer from vanishing and exploding gradients problem, which makes them difficult to train effectively, especially when dealing with long sequences of data. In simple terms, as the sequence length increases, the gradients calculated during back-propagation either shrink exponentially (vanish) or grow exponentially (explode), resulting in unstable learning.

\subsubsection{Sequential Computation}

RNNs process data sequentially, which prevents parallel computation within a sequence. This characteristic significantly slows down training and inference times, especially for long sequences.

\subsubsection{Difficulty in Capturing Long-Term Dependencies}

Although theoretically RNNs should be able to capture long-term dependencies due to their recurrent nature, in practice, they struggle to do so because of the vanishing gradient problem.

\subsection{How does PIPMN Handle Time-Series Data more Efficiently and Address Above Issues}

On the other hand, the PIPMN, despite not being a recurrent model, effectively handles time-series data, addressing the above issues.

\subsubsection{Temporal MLP and Depth Domain Block}

PIPMN incorporates Temporal MLP and Depth Domain Block to capture both spatial and temporal domain information. Temporal MLP can capture temporal dependencies across different time steps, and Depth Domain Block is used to extract the depth domain information more effectively.

\paragraph{Temporal MLP: Capturing Global Context Awareness}

The Temporal Multi-Layer Perceptron (TMLP) is responsible for understanding and extracting temporal dependencies across different time steps in audio data. Unlike the spatial domain which covers the breadth and depth of the sound, the temporal domain refers to the changes and patterns of sounds over time. Moreover, unlike RNNs, which process the input sequentially and focus on local context, Temporal MLP processes all time steps concurrently. This approach provides the model with a global view of the entire input sequence, enabling it to capture both short-term and long-term dependencies more effectively. The awareness of the global context allows the model to identify and understand patterns and relationships that span across a wide range of time steps.

\paragraph{Depth Domain Block: Handling Varied Depth Domain Information}

Depth Domain Block, on the other hand, is specifically designed to deal with the depth domain of audio data. The depth domain refers to the varying intensity or amplitude of sounds across different frequencies. Capturing depth domain information allows the model to distinguish sounds that have similar patterns but vary in intensity. By leveraging Depth Domain Block, PIPMN can extract depth information more effectively, thereby understanding the nuances of different sound amplitudes. This is essential in scenarios where similar sounds occur at different volumes or intensities, such as distinguishing between a shout and a whisper. On the other hand, RNNs are not designed to handle depth domain information, which can be a significant disadvantage in certain audio classification tasks.

\subsubsection{Avoids Sequential Computation}

Unlike RNNs, which process data in a sequential manner (input at time t depends on inputs at times t-1, t-2, and so on), Dense MLP blocks process all time steps concurrently. This means that the Dense MLP block can perceive the entire sequence at once, and the output at each time step is generated based on the full sequence. This ability to consider all time-steps at once can enable the model to capture both local and global temporal dependencies in the audio data.

\subsubsection{Handles Long-Term Dependencies Via PIPS}

The Paired Inverse Pyramid Structure in PIPMN facilitates the model to capture long-term dependencies without suffering from the issues faced by RNNs. The structure enables the model to extract and utilize information from different levels of abstraction, giving it the ability to capture both short and long-term dependencies more effectively.

In summary, our proposed PIPMN model is designed to address the challenges faced by RNNs in processing time-series data while enabling the model to capture both local and global temporal dependencies effectively. The model uses a combination of Temporal MLP and Depth Domain Block to extract essential features from audio signals, making it well-suited for audio classification tasks.

\subsection{Experiment Results Justification}

Based on the experimental results conducted on both the UrbanSound8K and Freesound2019 datasets, the choice of not using Recurrent Neural Networks (RNN), specifically the Long Short-Term Memory (LSTM) variant, was justified. When we replaced the DenseMLP component in our model with LSTM, we observed a noticeable drop in performance across various evaluation metrics. This indicated that, for this specific task, DenseMLP outperformed LSTM. The DenseMLP's superior performance could be attributed to its ability to extract a higher level of abstraction from the features, effectively dealing with both spatial and temporal dimensions of the data, which is crucial for sound classification tasks. In contrast, while LSTM is a powerful tool for handling sequential data, it may not be as efficient in modeling dependencies within the audio data or extracting relevant features as effectively as DenseMLP, leading to inferior performance in these particular experiments. These results reinforce the decision to opt for DenseMLP over LSTM in our model architecture for sound classification tasks.

\section{Conclusion}
\label{sec:typestyle}

In this paper, we propose a new network called PIPMN. The proposal of this network is based on the lightweight nature of audio. The results have shown that treating the audio spectrogram without adding a new dimension by virtual of the proposed Dense MLP block and Paired Inverse Pyramid Structure can achieve similar or even higher performance with much fewer parameters than those high-performing classification models. However, the PIPMN only focuses on ESC and MGC. For future work, we will extend our model for more challenging tasks such as sound detection.

% \section{Data Availability}
% All data included in this study are available upon request by
% contact with the corresponding author.
% UrbanSound8K dataset are available on http://urbansounddataset.weebly.com and GTAZN dataset are available on
% https://www.kaggle.com/
% \\
% datasets/andradaolteanu/gtzan-dataset-music-genre-classifica
% \\
% tion

% \section{Conflicts of Interest}
% The authors declare that they have no conflicts of interest
% regarding this work


\end{document}